\documentclass[a4paper,11pt]{article}
\pdfoutput=1 
\usepackage{jcappub} 
\usepackage{amssymb}
\usepackage{caption}
\usepackage{subcaption}
\usepackage{graphicx}
\usepackage{hyperref}
\usepackage[T1]{fontenc} 

\usepackage{cleveref}
\title{\boldmath  Probing memory-burdened Primordial Black Holes with global 21 cm signal}

\def\Tcmb{ {T_{\textrm{CMB}}} }

\begin{document}
\author[a,b]{Priyanka Sarmah}
\emailAdd{sarmahpriyanka07@gmail.com}
\emailAdd{sarmahpriyanka@gapp.nthu.edu.tw}

\author[a,b,c]{Kingman Cheung}
\emailAdd{cheung@phys.nthu.edu.tw}

\affiliation[a]{Department of Physics, National Tsing Hua University, Hsinchu 30013, Taiwan}
\affiliation[b]{Center for Theory and Computation, National Tsing Hua University, Hsinchu 30013, Taiwan}
\affiliation[c]{Division of Quantum Phases and Devices, School of Physics, Konkuk University, Seoul 143-701, Republic of Korea}

\date{\today}


\abstract{We investigate the imprints of memory-burdened primordial black holes (PBH) on the global 21 cm signal during the cosmic dawn. Recent studies reopened the possibility of a mass window of PBHs as a compelling candidate for dark matter,  particularly in low-mass regimes ($M_{\text {PBH}}< 10^{15}$ g) where conventional constraints from evaporation are being revisited in light of quantum gravitational effects. One such effect, the \textit{memory burden effect}, slows down black hole evaporation by incorporating the backreaction of radiation on the black hole microstates, substantially extending the lifetime of light PBHs and thus
modifying their late-time emission spectra. This prolonged emission can dramatically alter the energy injection history in the early universe.  By computing the modified energy injection rates into the intergalactic medium and incorporating them into the thermal and ionization evolution of neutral hydrogen, we obtain projected constraints on the fraction of dark matter. The bounds are obtained from the fact that these low mass PBHs, which were thought otherwise evaporated, can modify the absorption amplitude in the global 21-cm signal at redshift $z\approx17$. Considering the two viable scenarios of transition to the memory-burden phase: fast (or instantaneous) and slow (transition with a finite width), we show how the 21 cm bounds are sensitive to different mass ranges. For a broad transition with $\delta=10^{-2}$ we find that PBHs in the mass range $M_{\rm PBH}\simeq10^{8}$--$10^{13}$\,g  are excluded at the level of $f_{\rm PBH}\gtrsim10^{-8}$. In contrast, for a fast-transition case with the lowest suppression exponent $k=1$, the evaporation is suppressed so efficiently that no meaningful 21\,cm constraint remains for $M_{\rm PBH}\gtrsim10^{7}$\,g. }

\maketitle
\flushbottom

\section{Introduction}

Primordial black holes (PBHs) are considered to be one of the most motivated macroscopic dark matter (DM) candidates. Depending on their mass and production history, they may  account for the total or a fraction $f_{\rm PBH}$ of the 
whole cold dark matter abundance of the Universe\cite{Hawking:1971ei,Chapline:1975ojl,Khlopov:2008qy,Carr:2016drx,Carr:2020gox,Carr:2020xqk,Green:2020jor}.
A variety of mechanisms in the early universe can produce PBHs, including: (i) collapse of over-dense regions developed 
from primordial fluctuations after inflation~\cite{Carr:1974nx,Sasaki:2018dmp,Cheung:2023ihl},
(ii) bubble wall collisions with sufficient energy density within the Schwarzschild radius in first-order phase transitions~\cite{Hawking:1982ga,Kodama:1982sf,Moss:1994iq,Konoplich:1999qq}, 
and (iii) from the collapse of the macroscopic intermediate states called fermi-balls in the dark sector
\cite{Baker:2021nyl,Gross:2021qgx,Kawana:2021tde,Marfatia:2021hcp}. 

In the standard semiclassical picture, PBHs radiate thermally via Hawking evaporation. This process of evaporation is rapid   for light PBHs with masses 
below $10^{15}$ g~\cite{Hawking:1975vcx,Gibbons:1977mu}, and therefore only those above 
$M_{\rm PBH} \sim 10^{15}$--$10^{17}$ g can survive until today.
The standard Hawking evaporation is mainly controlled by the mass that determines the characteristic 
temperature $T_H=\frac{\hbar c^3}{8\pi G k_B M_{\text {PBH}}}=
1.06\frac{10^{13}\text{g}}{M_{\text{PBH}}}\,\text{GeV}$ and 
the timescale of evaporation $\tau_{\rm evap} = \frac{M_{\rm PBH}^3}{3\alpha} \simeq 
6.24\times 10^{-27}
\left( \frac{M_{\rm PBH} }{\rm g} \right)^3
{\rm s}$. 
Furthermore, the emitted particle spectra depend only on the PBH temperature and 
the masses and spins of the particles, regardless of their interaction strengths. For PBHs heavier than $M_{\rm PBH} \ge 10^{15}$g, numerous astrophysical and cosmological observations constrain their abundances $f_{\rm PBH}$. Such observations include microlensing
 \cite{Lensing1_Niikura:2017zjd, Lensing2_Niikura:2019kqi}, 
cosmic microwave background (CMB) anisotropies \cite{cmb1_Acharya:2020jbv, cmb2_Chluba:2020oip}, 
Big-Bang nucleosynthesis (BBN) \cite{bbn1_Carr:2009jm}, 
and gamma-ray observations \cite{bbn1_Carr:2009jm, gray1_Carr:2016hva}. 
In contrast, if Hawking radiation is valid throughout, PBHs with $M_{\rm PBH} \le 10^{15}$g cannot constitute the dark matter of the present universe, as they would have long gone due to the accelerated evaporation rate. Their evaporation can profoundly impact the thermal and ionization history of the early universe.
Moreover, these low mass PBHs may play a key role in processes such as 
reheating \cite{reheat1_He:2022wwy, reheat2_RiajulHaque:2023cqe}, 
leptogenesis \cite{lepto1_Calabrese:2023key, lepto2_Gunn:2024xaq}, and 
production of gravitational waves \cite{gw1_Inomata:2019ivs, gw2_Sugiyama:2020roc, gw3_Inomata:2020lmk, gw4_Papanikolaou:2020qtd, gw5_Domenech:2021wkk, gw6_Papanikolaou:2022chm, gw7_Balaji:2024hpu, gw8_Bhaumik:2022zdd}, in cosmic acceleration of universe driving inflation\cite{lightpbh1Dialektopoulos:2025mfz}, also in alleviating the $H_0$ tension \cite{lightpbh2Papanikolaou:2023oxq}. 
Their evaporation can also affect the redshifted 21-cm signal, arising from the neutral hydrogen, which was present in abundance during the cosmic dark ages,  through excessive heating of the intergalactic medium (IGM)\cite{Mittal:2021egv, Saha:2021pqf, Natwariya:2021xki}.
A recently claimed detection by the EDGES experiment reported an absorption signal 
at $z \sim 17$ with $T_{21} \approx -500$ mK, significantly deeper than standard predictions, though this result remains debatable \cite{Bowman2018}. Regardless of its validity, conservative upper limits on the magnitude of absorption or emission of 21-cm photons at high redshifts can translate into bounds on excessive 
heating sources, such as the bounds on PBH abundance and properties. This is the key logic behind this work.

Hawking radiation is based on the semi-classical treatment of black hole (BH) dynamics.
Therefore, the energy spectrum of the particles coming off the PBHs with mass $M_{\text {PBH}} \sim 10^{14-15}\,$g 
peaks at about 10-100 MeV, as the Hawking temperature is inversely proportional to their mass. 
Nevertheless, recent works \cite{Dvali:2018xpy,Dvali:2020wft,Dvali:2024hsb}  have explored how quantum effects may alter this picture. In particular, they argued that including the backreaction of emission on the internal quantum state of the black hole may modify the standard scenario of Hawking
radiation, by suppressing the further emission. This particular phenomenon is coined as {\it memory burden effect} - a quantum-gravitational effect
proposed to regulate the late-time evaporation of black holes.
It is caused by the information stored in the BH, which resists its evaporation. 
Therefore, once the BH loses a certain fraction of
its initial mass, the backreaction becomes significant enough to suppress or slow down the
evaporation process, thereby extending the BH lifetime. Due to the memory burden (MB) effect,
for example, $10^6 - 10^{14}$g PBH can still survive to this day. For such lighter PBHs, the 
energy spectrum for the emission of particles would shift to higher energies, and thus modify 
the energy injection to the early Universe. It would give different effects to, e.g., CMB, Lyman-alpha,
21-cm radiation, etc. Several recent works have developed the phenomenology of 
memory burdened PBHs and explored their observational consequences across cosmology and astrophysics. 
Early studies of the MB effect originated from the framework of Refs.~\cite{Dvali:2018xpy,Dvali:2020wft,
Dvali:2024hsb}, which showed that the backreaction of the emitted quanta on the initial quantum state of 
the black hole leads to a suppression of the Hawking radiation after a critical fractional mass loss. 
Building on this idea, a number of works dedicated to \emph{instantaneous} (or fast) transition to 
the MB regime, in which the PBH evaporation rate drops abruptly once the transition point is reached \cite{Ft1_Alexandre:2024nuo, Ft2_Loc:2024qbz, Ft3_Athron:2024fcj, Ft4_Zantedeschi:2024ram, Ft5_Thoss:2024hsr, Ft6_Chaudhuri:2025rcs, Ft7_Borah:2024bcr, Ft8_Barman:2024ufm, Ft9_Bhaumik:2024qzd, Ft9a_Chianese:2024rsn, Ft9b_Calabrese:2025sfh, Ft10a_Chianese:2025wrk, Ft10_Tan:2025vxp, Ft11_Chaudhuri:2025asm}.  Ref.~\cite{MBbeyondpbbh_Basumatary:2024uwo} studies the memory-burden effect in the context of particle DM accretion-seeded black holes within neutron stars.

On the other hand, Refs.~\cite{smoothtrans2_Montefalcone:2025akm, smoothtrans1_Dvali:2025ktz} pointed out that realistic transitions are expected to occur gradually, with the evaporation rate suppressed continuously over a finite range of mass loss. This led to studies analyzing slow or continuous transition, which yield stronger constraints.  Ref.~\cite{fastnslow} performed a comprehensive treatment of fast, slow, and merging MB transitions, combining BBN, CMB anisotropies, and diffuse $\gamma$--ray data.  The slow transition case leads to substantial late-time energy injection that can be tightly constrained by late-time cosmological observables like BBN, the CMB, and 
high-energy photon and neutrino backgrounds. This motivates the use of another complementary probe of the physics of the late-time universe: the hydrogen 21-cm line. The global 21-cm signal, which is particularly sensitive to small amounts of energy injection at $z\sim10$--20, can offer a sharp discriminator between fast and slow memory burden transition scenarios.

In this work, we focus on the effect on the 21-cm  signal in the presence of MB effect
for PBHs of mass below $10^{15}$g. We calculate energy emission from PBHs and how it affects 
the thermal history of baryonic matter (gas temperature $T_m$), and 
thus changing the 21-cm brightness temperature contrast $ T_{\rm 21}$. 
We estimate the sensitivity of 21-cm signal observation on the parameter of PBHs. 
The organization of this work is as follows. In \cref{sec:MB_pheno}, we review the impact of memory burden on the lifetime and emission energy spectrum of PBHs. We discuss the two possible scenarios of memory burden scenarios-- the instantaneous (fast) and gradual (slow) transitions.
In \cref{sec:21cm_details}, we outline the physics of the 
global 21-cm signal and how it is modified by PBH heating. 
We present numerical results and derive constraints on the memory burden parameters corresponding to the scenarios of fast and slow transition and
parameters of PBHs $(M_{\text{PBH}}, f_{\text{PBH}})$. We summarize our findings in \cref{sec:discussion} with a discussion on existing constraints from CMB and BBN.


\section{Phenomenology of Memory Burden effect}\label{sec:MB_pheno}

The essence of the memory burden (MB) effect stems from the fact that a system with large microstate degeneracy wants to stabilize its high capacity of information, such as black holes or other objects referred to as “saturons" \cite{Dvali:2024hsb}. In this framework, the emission of quanta perturbs the internal microstate structure, producing a quantum backreaction that raises the effective energy threshold for subsequent emission. As a result, after a black hole has shed a certain fraction of its initial mass, further evaporation becomes increasingly suppressed. 
For PBHs this mechanism can lengthen its lifetime. 
Consequently, light PBHs that would otherwise have evaporated are expected to be 
present today as viable dark matter candidates, reopening the mass window below  $10^{15}\,\mathrm{g}$.

In the semi-classical picture of Hawking radiation, a non-rotating black hole of mass $M_{\rm PBH}$ 
radiates thermally at a characteristic temperature
\begin{equation}
T_H=\frac{\hbar c^3}{8\pi G k_B M_{\rm PBH}}=1.06\frac{10^{13}\text{g}}{M_{\rm PBH}}\;\text{GeV}\,.
\end{equation}
In natural units ($\hbar=c=k_B=1$) one often uses $T_H = 1/(8\pi G M)\propto 1/M_{\rm PBH}$.
The BH carries an entropy given by the Bekenstein--Hawking entropy
\begin{equation}
\label{eqn:entropy}
S(M)=2\pi M r_g\;\simeq\; 10^{30} \times 
\left( \frac{M_{\rm PBH}}{10^{10}\,\mathrm{g}} \right)^2\,,
\end{equation}
where $r_g = 2GM/c^2 \sim M $ is the Schwarzschild radius of the BH. 
The semiclassical mass-loss rate and lifetime scale as (order-of-magnitude) 
  \begin{equation}
\left(\frac{dM_{\rm PBH}}{dt}\right)_{\rm sc}\sim -\frac{1}{r_g^2}\sim -\frac{1}{M_{\rm PBH}^2},
  \qquad \tau_{\rm sc}\sim S\,r_g, \sim M_{\rm PBH}^3,
   \end{equation}

Hence, in the semi-classical picture, the lighter the BH mass, the faster the radiation rate and the shorter the lifetime 
of the BH, implying that  PBHs of mass less than $10^{15}$g had already evaporated
by now. The differential particle emission rate from such an evaporating non-rotating BH with mass M$_{\rm PBH}$ is given by
\begin{equation} \label{eqn: dndedt_semicalss}
\frac{d^2N_{i}^{sc}}{dE \, dt}(E, M_{\rm PBH})= \frac{g_i}{2\pi}
    \frac{\Gamma_i(E,M_{\rm PBH})}{\exp(E/k_BT_{\rm H}) \pm 1},
 \end{equation}
where $i \in \{\gamma, e^\pm, \nu, p, \bar{p}, \ldots\}$ labels the 
particle species. $g_i$ is the degrees of freedom for particle species $i$ and $\Gamma_i$ 
are the greybody factors, and $\pm$ corresponds to the particle $i$ being a fermion or a boson,
respectively. 

Nevertheless, when the MB effect sets in, the semi-classical radiation rate is no longer valid. 
In the simplest implementation, the MB sets in when the BH loses a fraction $q$ of its original
mass. Such an instantaneous onset often chooses $q=1/2$ in the literature \cite{Ft1_Alexandre:2024nuo}. It means that after 
the BH lost one-half of its original mass according to the semi-classical approximation; it would be
followed by the MB phase, which suppresses the evaporation rate, conventionally given as,
\[
  \left(\frac{dM}{dt}\right)_{\rm MB}\sim -S^{-k}\left(\frac{dM}{dt}\right)_{\rm sc}\quad\Rightarrow\quad \tau_{\rm MB}\sim S^{1+k}r_g .
\]
where $\tau_{\rm MB}$ is the lifetime in the MB phase. 
Here  $k$ is expected to be an integer ($k=1,2, ...$) characterizing the MB suppression strength. Because $S$ is a large number, even $k=1$ leads to highly suppressed emission rate, 
and even more suppressed as $k$ increases. This elongates the lifetime of the black hole given by the approximate form ~\cite{Dvali:2018xpy,Dvali:2020wft,Dvali:2024hsb}
\begin{equation}
	\tau_{\rm MB} \simeq S^{\,1+k} \, r_g \,.
\end{equation}

The benchmark redshift for the global 21-cm signal that we are interested in is $z\approx 20$. 
Setting $\tau = t(z=20)$ yields $M_{\rm PBH} \simeq 4\times10^{4}\,$g, $3\times10^{-2}\,$g, and $6\times10^{-6}\,$g for $k=1,2,3$, respectively. In our case, we treat $k$ as a free  parameter. 
The value $k=2$ is adopted as a theoretically motivated benchmark in the memory-burden framework. This choice is motivated by entropy and information-theoretic arguments, according to which the information stored in black-hole microstates backreacts on the late-time evaporation process and suppresses the Hawking emission rate by inverse powers of the black-hole entropy. In this parametrization, $k=2$ corresponds to a quadratic inverse-entropy suppression and is therefore commonly used as a motivated benchmark \cite{Dvali:2020wft, Dvali:2024hsb, kmotivation2Dvali:2017nis}. In contrast, other values of $k$ considered in this work are treated as phenomenological variations, introduced to explore the sensitivity of our results to the strength of the suppression.

Several recent works have emphasized that the onset of the MB regime may occur gradually rather 
than abruptly \cite{smoothtrans2_Montefalcone:2025akm,smoothtrans1_Dvali:2025ktz,fastnslow}. This motivates the introduction of a 
transition-width parameter $\delta \ll 1$, which encodes the finite interval of mass loss over which the backreaction becomes significant.  The MB effect sets in gradually over a finite mass fraction 
$\delta\ll q$.  In our analysis, we adopt $q = 1/2$ as a benchmark value that controls the onset of the memory-burden backreaction. A microscopic Hamiltonian analysis \cite{smoothtrans1_Dvali:2025ktz}
showed that the transition width $\delta$ satisfies
\begin{equation}
\label{q-delta}
\delta \gtrsim \frac{q}{(p-1)\ln S},
\end{equation}
where $S$ is the black hole entropy and $p$ is a positive integer
characterizing the growth rate of the backreaction.
This relation arises from the fact that $\delta$ is set by how
rapidly the memory burden backreaction grows as the black hole departs
from the critical point of gaplessness, which is governed by the
exponent $p$ and the PBH entropy $S$.
For the well-motivated case $p=2$, this relation implies
$\delta \sim 10^{-2} q$, while even for $p \gg 2$ one expects
$\delta \gtrsim 10^{-4} q$.
This indicates that $q$ and $\delta$ are not generically independent,
but instead scale in a correlated manner,
with larger values of $q$ corresponding to slower transitions
(larger $\delta$), and vice versa.
Nevertheless, since a complete theoretical description of the MB transition is still lacking, it is useful to explore $\delta$ phenomenologically at fixed benchmark $q$. Including this possibility of a slow transition, the suppression factor for times $t\gtrsim \tau_{\rm sc}/2$ scales as \cite{fastnslow}
\begin{equation}
\label{eq:suppression_factor}
    \mathcal{K}(t, M_{\rm{PBH}}) \simeq \max\!\big(S^{-k},\, \delta\, \tau_{\rm sc}/(2t)\big)
\end{equation}
with the two limiting regimes:
\begin{itemize}
  \item $S^{-k}$ — instantaneous transition,
  \item $\delta \,\tau_{\rm sc}/(2t)$ — slow transition
\end{itemize}
We study these two limiting cases, considering $k$ and $\delta$ to be free parameters, as there is no theoretical relation between them. In this work, we scan the PBHs with mass range $M_{\rm{PBH}}=10^3 -10^{14}$ g. The instantaneous-transition limit is recovered for $\delta \to 0$, while $\delta > 0$ produces a finite transition epoch during which the evaporation rate evolves smoothly from the semiclassical rate to the MB-suppressed regime.

\begin{figure}
    \centering
    \includegraphics[width=0.48\linewidth]{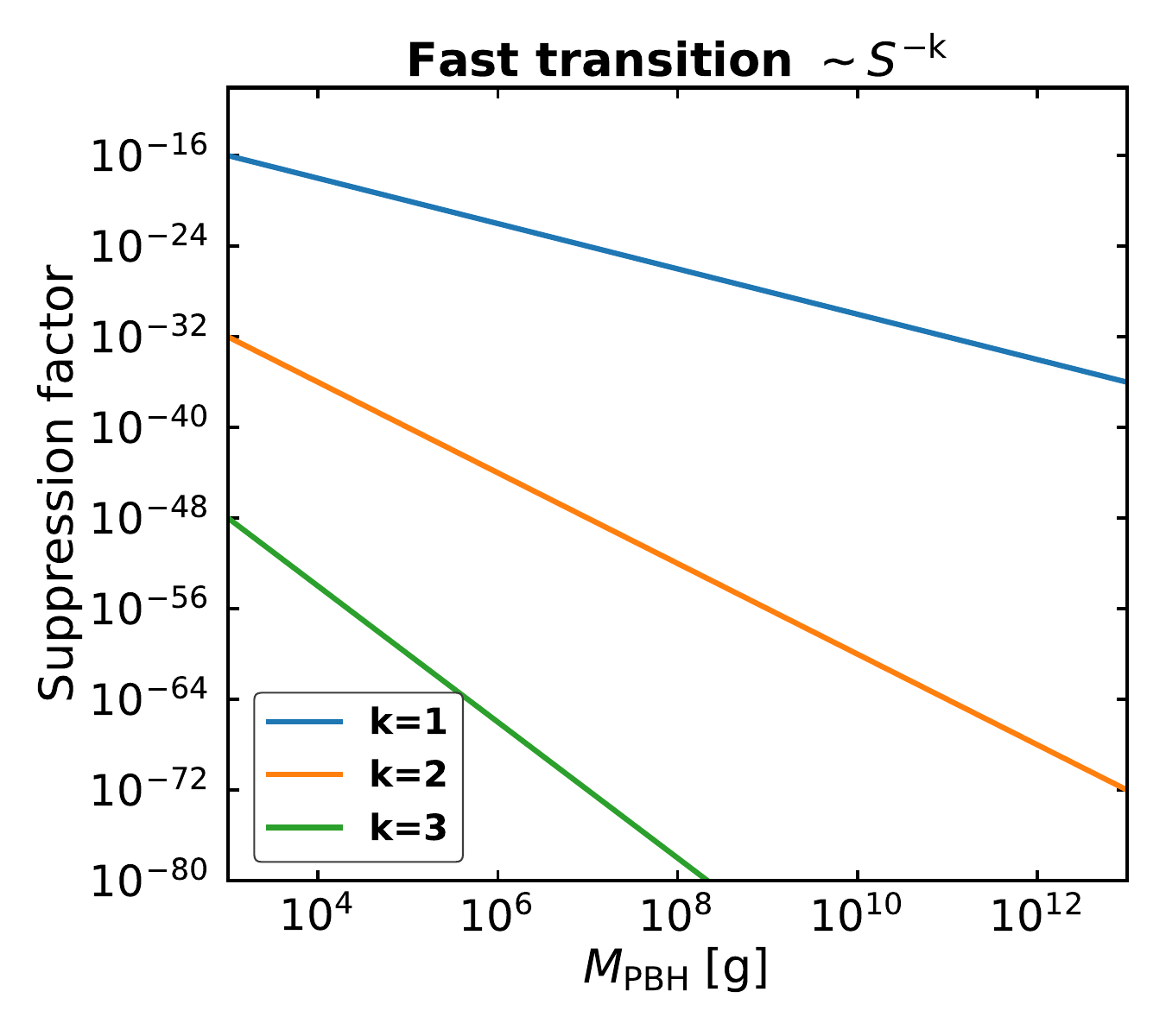}
    \includegraphics[width=0.48\linewidth]{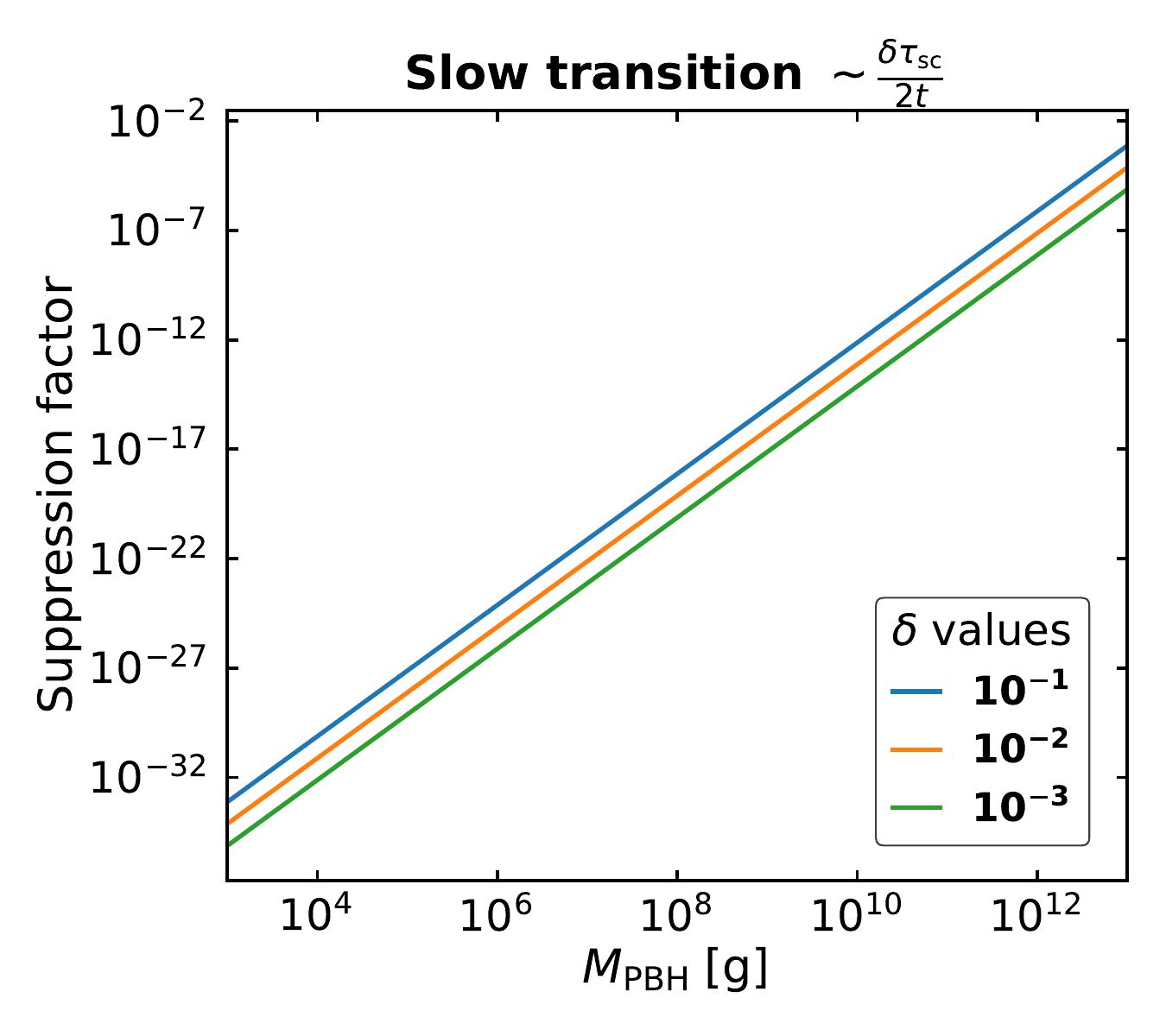}
    \caption{Suppression factor $\mathcal{K}$ as a function of PBH mass for the fast and slow memory-burden transitions. \emph{Left:} Fast-transition limit, where the evaporation rate changes instantaneously. In this regime, the suppression follows the entropy scaling $S\!\propto\! M^{2}$, giving $\mathcal{K}\!\propto\! M^{-2k}$. As a result, the curves (shown for $k=1,2,3$) become increasingly steep with larger~$k$, and higher-mass PBHs are suppressed much more strongly.
    \emph{Right:} Slow-transition regime ($\delta>0$), where the evaporation evolves smoothly over a finite interval. Here, the heavier PBHs experience \emph{less} suppression because their longer semiclassical lifetime $\tau_{\rm sc}\propto M^{3}$. Curves for different $\delta$ values show that a larger  $\delta$ produces broader and more gradual suppression.}
    \label{fig:supp_fact}
\end{figure}

The scaling of the suppression factor is illustrated in \cref{fig:supp_fact}. The fast transition regime, shown in the left panel, exhibits a scaling $\mathcal{K} \sim M^{-2k}$, arising from scaling of the entropy, $S \sim M^2$. As illustrated in the left panel of \cref{fig:supp_fact}, the suppression factor shows a steeper mass dependence: heavier PBHs experience stronger suppression, and the slope becomes even steeper for larger $k$. Thus, for fixed $M_{\rm PBH}$, increasing $k$ enhances the suppression, and at fixed $k$, the high-mass end of the spectrum is most strongly affected. 

On the other hand, the PBH does not transit instantaneously to the MB phase in the slow-transition regime, but rather transits with a finite width of the transition governed by the parameter ~$\delta$. 
The right panel of \cref{fig:supp_fact}, evaluated at a fixed cosmological time $t$ (corresponding to a 
redshift $z=20$), depicts that for a fixed $M_{\rm PBH}$, a larger transition width ($\delta=10^{-1}$) produces a  moderate suppression, whereas smaller widths such as  $\delta=10^{-2}$ and $\delta=10^{-3}$ yield progressively stronger suppression.
The mass dependence in the slow-transition case differs from the fast-transition case. 
The suppression factor $\mathcal{K}$ exhibits a characteristic mass dependence inherited from the classical evaporation timescale $\tau_{\rm sc} \propto M^3$. 
More massive PBHs enter the transition region later in time, implying that their evaporation will be less affected by the memory-burden effect. 
Lighter PBHs, in contrast, encounter the transition much earlier, resulting in a stronger 
net suppression. 
Hence, heavier PBHs encounter \emph{less} suppression in slow transition, which is in contrast with the entropy-driven $M^{-2k}$ scaling in the fast-transition regime.


Depending upon the nature of the transition, the modified spectra in the MB phase
is given by
\begin{equation}
\left.\frac{d^2N_i}{dE \, dt}\right|_{\rm MB}(E, t, M_{\rm PBH}) = \mathcal{K}(t, M_{\rm PBH}) \cdot \frac{d^2N_{i}^{sc}}{dE \, dt}(E, M_{\rm PBH}),
\end{equation}
where $\mathcal{K}(t, M_{\rm{PBH}})$  encodes the deviation from the semiclassical picture, as given in \cref{eq:suppression_factor}. We calculate the semiclassical emission rate in \cref{eqn: dndedt_semicalss} using the publicly available code \textbf{BlackHawk2.3} \cite{Arbey2021BlackHawk2}.

\section{21-cm line as a probe of matter temperature} \label{sec:21cm_details}
The cosmic 21-cm line originates from the hyperfine transition of neutral hydrogen. Any transition between the split states either emits or absorbs
a photon with a rest wavelength of 21 cm or equivalent frequency of 1420 MHz. During the cosmic dawn and reionization epochs (roughly $6 < z < 30$), this signal traces the thermal and ionization state of the intergalactic medium (IGM) through the differential brightness temperature $T_{21}$ relative to the cosmic microwave background. This observable is given by \cite{Pritchard2012}
\begin{equation}\label{eqn:T21}
	T_{21} \approx 23 \text{ mK} \; x_{HI}(z) \left(\frac{\Omega_b h^2}{0.02}\right) \left(\frac{0.15}{\Omega_m h^2}\right)^{1/2} \left(\frac{1+z}{10}\right)^{1/2} \left(1 - \frac{T_{\rm CMB}}{T_s}\right),
\end{equation}
where $T_{\rm CMB}$ is the CMB temperature at redshift $z$, $T_s$ is the spin temperature of the neutral  hydrogen, $\Omega_b$ and $\Omega_m$ are the baryon and matter density parameters, and $h$ is the dimensionless Hubble parameter. In our work, we adopt values
$\Omega_b h^2 = 0.0224$, $\Omega_m h^2 = 0.142$, and $h = 0.67$ \cite{Planck:2018vyg}. During the cosmic dawn ($z\sim1000$ to $z\sim 10$.), the neutral fraction $x_{HI}(z) $is close to unity,  such that global 21-cm observations serve as a sensitive probe of deviations in the thermal evolution of the intergalactic medium.
The spin-temperature $T_s$ is defined in terms of the relative populations of the triplet and singlet states, such that $n_1/n_0 = 3 e^{-\frac{\Delta}{T_s}}$, where $ \Delta = (h c)/( k_B \, 21 \, \textrm{cm}) =   69~\textrm{mK} = 5.9 \,\mu\textrm{eV} $ is a temperature/energy scale corresponding to the hyperfine-splitting and $n_1$ and $n_0$ are the number densities of the triplet and singlet states respectively. In standard cosmology,  the spin temperature is coupled to the CMB temperature $T_{\rm CMB}$ and to the matter temperature $T_m$ mainly through processes  such as gas collisions, and  to
Lyman-$\alpha$ photons from stars and galaxies, which depend on astrophysical modeling ~\cite{Furlanetto:2006tf}. 

\begin{equation} 
    T_s^{-1} = \frac{{\Tcmb}^{-1} + x_{C}T_{m}^{-1} + x_{\alpha} T_{c}^{-1} } {1 + x_C + x_{\alpha}}
    \label{eqn:Ts}
\end{equation}
where $T_c$ is the color temperature of Lyman-$\alpha$ photons from the first stars~\cite{Barkana:2004vb}. 
In this work, we assume no interaction with Lyman-$\alpha$ (thus $x_\alpha = 0$).

The strength of the coupling between $T_s$ and $T_m$ determined by the effective coupling $x_C$ known as the collisional coupling, arises because of the scattering of the hydrogen atoms with each other or with other species like electrons/protons \cite{Pritchard2012}. 
 Depending on the background temperature $T_{\rm CMB}$ and the spin temperature $T_s$, the 21-cm signal has features of (i) absorption ($T_{\rm CMB} > T_s$), (ii) emission ($T_{\rm CMB} < T_s$) and (iii) no signal ($T_{\rm CMB} = T_s$).
The standard cosmology predicts two distinct absorption dips in the global 21-cm brightness temperature, one around ($z\sim 46-140$), mainly controlled by the gas collision dynamics, and the other around ($z\sim 15$) due to Lyman-$\alpha$ photons from the first stars.

Energy injection from exotic sources, including evaporating PBHs, can significantly perturb the expected cooling history of gas, thus altering the evolution of $T_m$ (hence $T_s$) and the predicted 21-cm signal \cite{Mittal:2021egv, Saha:2021pqf, Natwariya:2021xki}.
In the standard $\Lambda$CDM cosmology without additional energy sources, the matter temperature declines adiabatically as $T_m \propto (1+z)^2$ after decoupling from the CMB. In contrast, the CMB temperature scales as $T_{\rm CMB} \propto (1+z)$. As a result, around $z \sim 200$, the matter temperature falls below the CMB temperature \cite{Pritchard2012}. However, Hawking radiation from PBHs injects high-energy photons and $e^\pm$ pairs, which trigger electromagnetic cascades and deposit a fraction of their energy into the IGM through
heating, ionization, and excitation of hydrogen and helium.

The energy injection rate density from PBHs into particle species $i$  at redshift $z$ is (\cite{PBHheating_PhysRevLett.121.011103, Natwariya:2021xki})
\begin{equation}
\frac{dE_{i}^{\rm PBH}}{dV \, dt}(z)\bigg|_{\rm inj} = n_{\rm PBH}(z) \int_0^\infty E \, \frac{d^2N_i}{dE \, dt}\bigg|_{\rm MB}(E, t(z), M_{\rm PBH}) \, dE,
\end{equation}
where the PBH number density is
\begin{equation}
n_{\rm PBH}(z) = \frac{f_{\rm PBH} \rho_{\rm DM,0}}{M_{\rm PBH}}(1+z)^3,
\end{equation}
with $\rho_{\rm DM,0} = \Omega_{\rm DM} \rho_{c,0}$ being the today's dark matter density,   and $\rho_{c,0} = 3H_0^2/(8\pi G)$ represents the critical density.

In the presence of the above energy injection, the matter temperature $T_m$ evolves according to (\cite{PBHheating_PhysRevLett.121.011103, Natwariya:2021xki})
\begin{equation}
\frac{dT_{ m}}{dz} = 
\frac{ 2 T_{ m} }{1+z}-
\frac{\Gamma_C}{1+z} \big( T_\gamma(z) - T_{\rm m} \big)
 - \frac{1}{(1+z) H(z)}
 \left( \frac{dE}{dV\, dt} \right)_{\rm inj}^{\rm PBH}
 \frac{1}{n_H}
 \frac{2 f_{\rm heat}(z)}{3 (1 + x_e + f_{\rm He})} ,
 \label{eq:Tm_evolution}
\end{equation}
where  $H(z) = H_0 \sqrt{\Omega_m (1+z)^3 + \Omega_\Lambda}$ is the Hubble expansion rate,  $T_\gamma = T_{\rm CMB,0}(1+z)$ is the CMB temperature, $x_e$ is the free electron fraction and $f_{\rm He}$ is helium number density. The first term in the evolution accounts for adiabatic cooling due to the universe's expansion, while the second term captures cooling/heating of the gas through Compton scattering, with $\Gamma_c$ standing for the Compton-scattering rate.

Light PBHs ($M_{\rm PBH} \lesssim 10^{17}$\,g)  predominantly emit high-energy photons and electron-positron pairs that initiate electromagnetic cascades, thermalizing a fraction of the injected energy as heat. The heating efficiency depends on the redshift and the spectrum of injected particles. In this work, we use the \textbf{DarkHistory} package \cite{Liu:2019bbm} to deal with the matter temperature evolution in the presence of  PBH-induced energy injection. Following the method outlined in \cite{Saha:2024ies},  we modify the injection module to incorporate the injected spectra from \textbf{BlackHawk}, including the backreaction of energy deposition. 

For our analysis, we focus on $z=17$ as a fiducial redshift where 21-cm observations provide stringent constraints. At this epoch, the standard adiabatic cooling predicts $T_m \approx 6.8$ K (significantly colder than the CMB temperature $T_{\rm CMB} \approx 49$ K), yielding a moderately strong absorption signal. We define an upper bound on the matter temperature $T_{\rm max}$ corresponding to a given 21-cm brightness temperature threshold. For example, assuming that the signal does not exceed $T_{21} = -150$ mK translates via the \cref{eqn:T21} to $T_{\rm max} \approx 9.41$ K at $z=17$. 
For our analysis, we adopt $z=17$ as a fiducial redshift near the expected epoch of peak 21-cm absorption during cosmic dawn.
\footnote{The choice of $z=17$ as a fiducial constraining epoch is motivated by the location of the global 21-cm absorption feature reported by the EDGES collaboration, which claimed an absorption profile centered around $\nu\sim 78$~MHz $(z\sim 17)$ \cite{Bowman2018}. The interpretation of this result remains debated: SARAS~3 reported a non-detection of a signal with the EDGES amplitude at 95\% confidence \cite{Singh:2021mxo}, and several works have raised concerns about the foreground modelling and sky-averaged analysis methods (e.g.\cite{Hills:2018vyr, Sims:2019kro}). Still, a cosmological origin remains possible and motivates studies of non-standard thermal histories. For the injection histories considered here, including the $\sim t^{-1}$ scaling, the cumulative heating is strongly constrained near this epoch because the gas is cold relative to the CMB. We have verified that evaluating the constraint over $z\sim 15$--$20$ gives comparable bounds.}
At this epoch, standard adiabatic cooling predicts $T_m \approx 6.8$ K, which is significantly lower than the CMB temperature $T_{\rm CMB} \approx 49$ K, leading to an absorption signal. To quantify the impact of additional heating, we introduce a benchmark upper bound on the matter temperature $T_{\rm max}$ derived from a chosen threshold on the 21-cm brightness temperature. In particular, requiring $T_{21} \ge -150$ mK corresponds, via \cref{eqn:T21}, to $T_{\rm max} \approx 9.41$ K at $z=17$. This choice is adopted as a conservative benchmark motivated by typical $\Lambda$CDM expectations \cite{lambdacdmReis:2021nqf}, rather than as a firm observational bound.
Any PBH scenarios predicting  $T_m > T_{\rm max}$ are thus disfavored by this constraint. 
Under the assumptions considered here, this approach allows the 21-cm observable to provide projected sensitivity to memory-burdened PBH scenarios, since different values of $\delta$ and $k$ lead to different suppression factors and, consequently, different heating rates.
\begin{figure}
    \centering
    \includegraphics[width=0.48\linewidth]{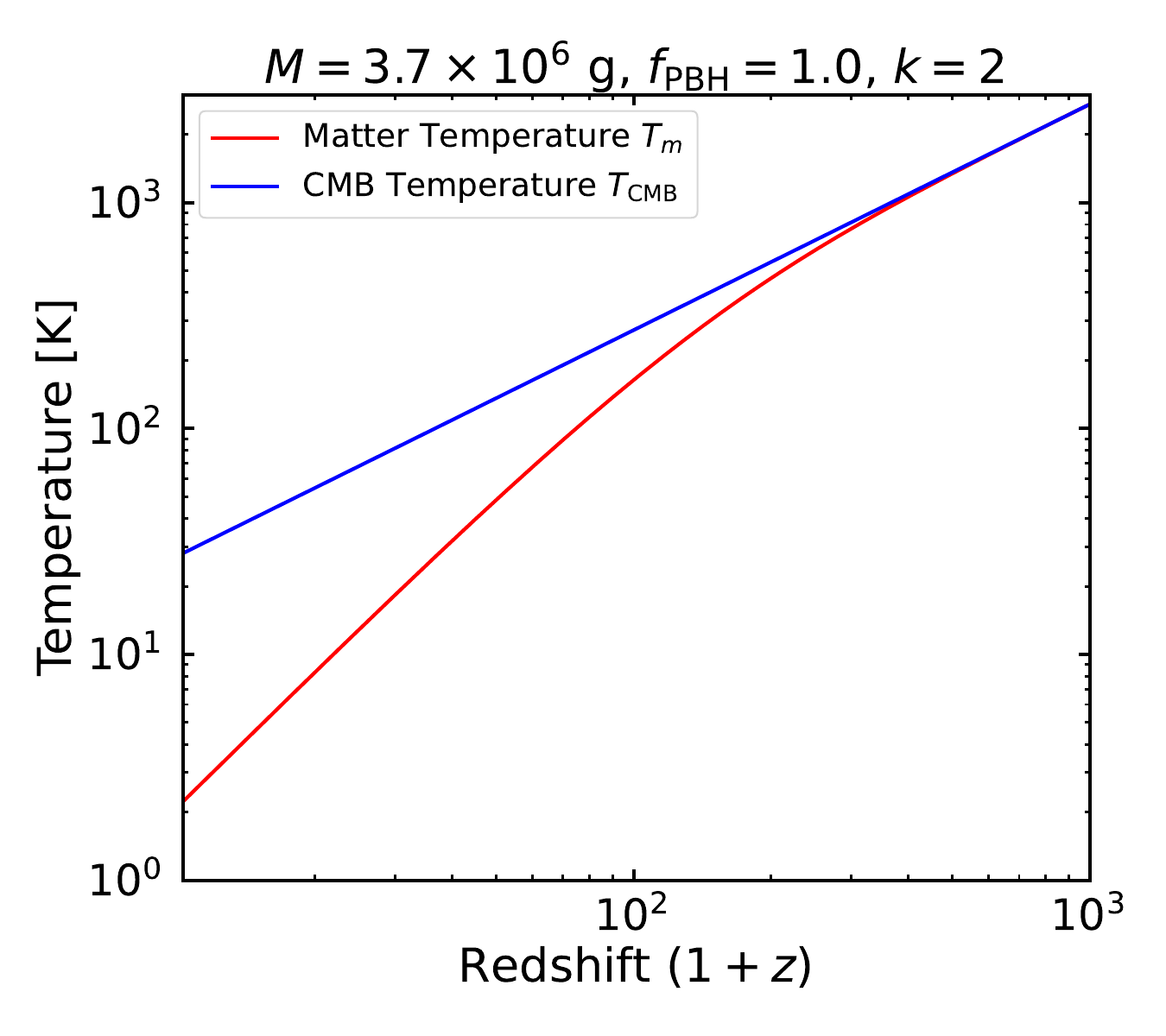}
    \includegraphics[width=0.48\linewidth]{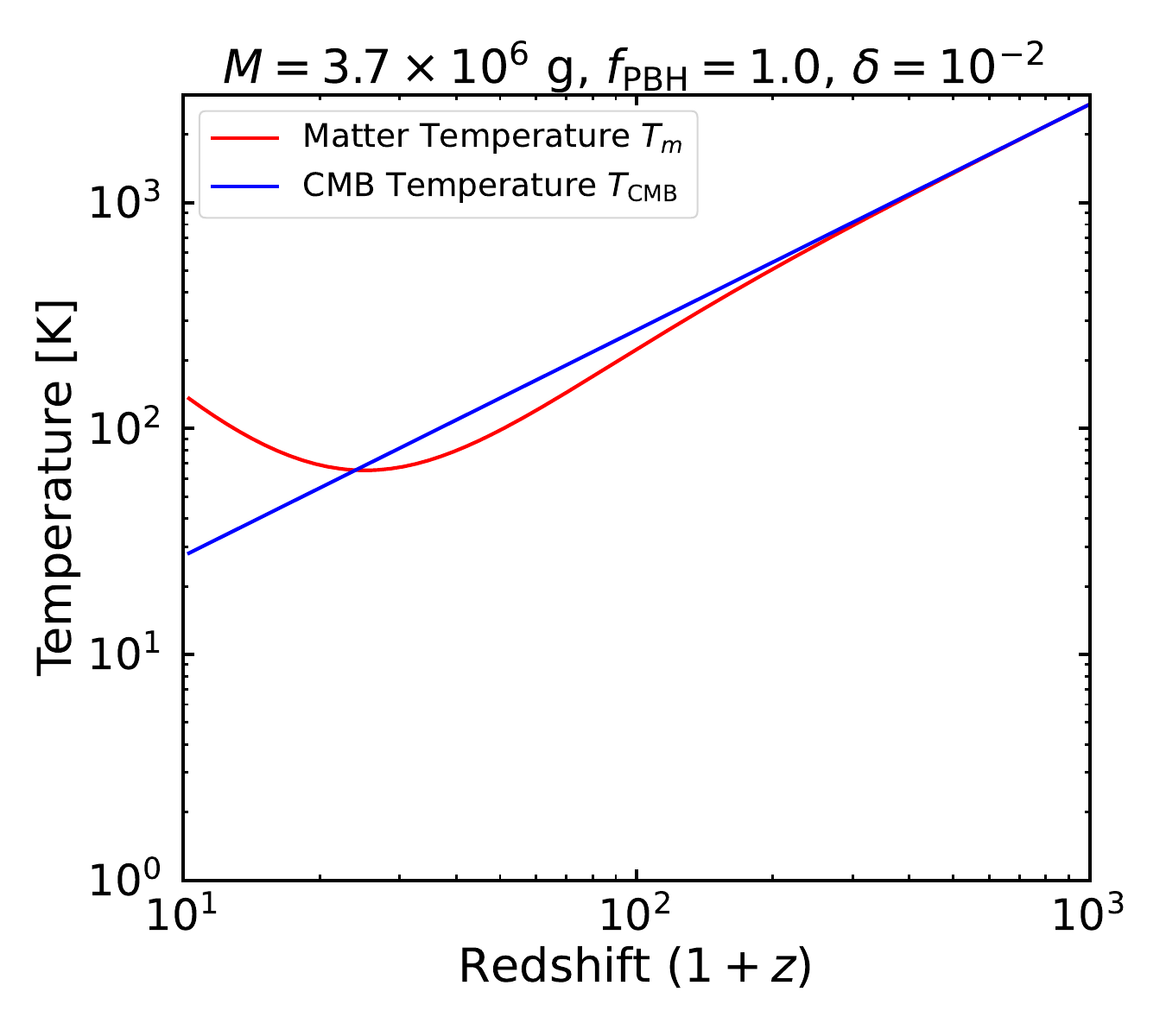}
    \caption{Redshift evolution of the matter temperature  $T_m$ for a benchmark PBH mass 
	$M_{\rm PBH}=3.7\times10^{6}\,\mathrm{g}$ and $f_{\rm PBH}=1$, comparing fast (left) and slow (right) transition to memory burden regime. A fast transition ($k=2$) produces an abrupt suppression of Hawking radiation through $S^{-k}$ scaling, leading to only modest heating and $T_m$ stays 
	below $T_{\rm CMB}$ for all relevant redshifts. For the slow-transition case ($\delta = 10^{-2}$), the suppression turns on gradually, and the PBH continues to evaporate at an appreciable fraction of its semiclassical rate. This prolonged emission injects substantial energy into the IGM, raising $T_m$ above $T_{\rm CMB}$ around $1+z\approx25$. }
    \label{fig:Tm_evolution}
\end{figure}

We show the redshift evolution of the matter temperature $T_m$ in \cref{fig:Tm_evolution} for a representative primordial black hole mass $M_{\rm PBH} = 3.7\times10^{6}\,\mathrm{g}$ and $f_{\rm PBH}=1$, illustrating how the thermal history of the IGM depends on the transition dynamics of the memory burden effect.

For the fast transition scenario (left panel), shown for $k=2$, the onset of the memory burden regime is instantaneous, and the emission is strongly suppressed by the $S^{-k}$. Hence, after a brief initial emission, due to largely suppressed spectra, the subsequent energy injection into the IGM is minimal. As a result, the matter temperature $T_m$ closely follows the adiabatic cooling trajectory,  remaining consistently well below the CMB temperature. This case, therefore, produces no significant heating feature in the 21 cm signal. In contrast, in the slow-transition scenario (right panel, $\delta = 10^{-2}$), the suppression of Hawking emission develops only gradually with non-zero transition width $\delta$. As a result, the PBHs continue to radiate at an appreciable fraction of their semiclassical rate over an extended period. This sustained injection of energetic photons and $e^\pm$ pairs heats the IGM, raising $T_m$ above the standard adiabatic cooling and therefore above background temperature $T_{\rm CMB}$ at a redshift of $1+z\approx 25$. This results in a reduced amplitude of the 21 cm absorption signal. 

Thus, for the same $(M_{\rm PBH}, f_{\rm PBH})$ benchmark, the slow transition with a non-zero width to memory-burden phase can lead to a thermal excess in the baryonic gas, whereas the fast transition fails to generate such departure from the CMB temperature, highlighting the role of the transition timescale in determining the observable impact of the memory-burdened PBH evaporation on 21 cm line.

\begin{figure}
    \centering
    \includegraphics[width=0.49\linewidth]{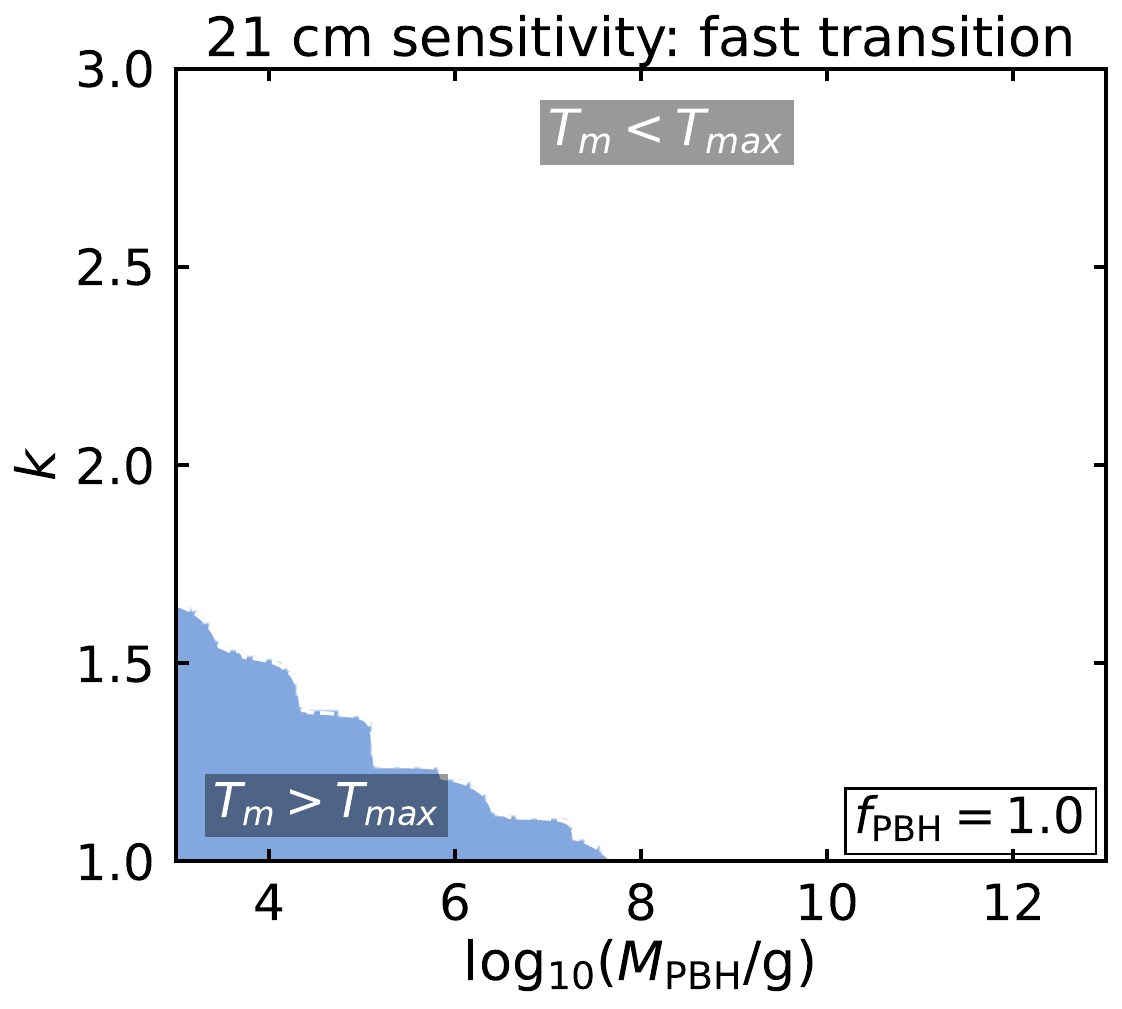}
    \includegraphics[width=0.49\linewidth]{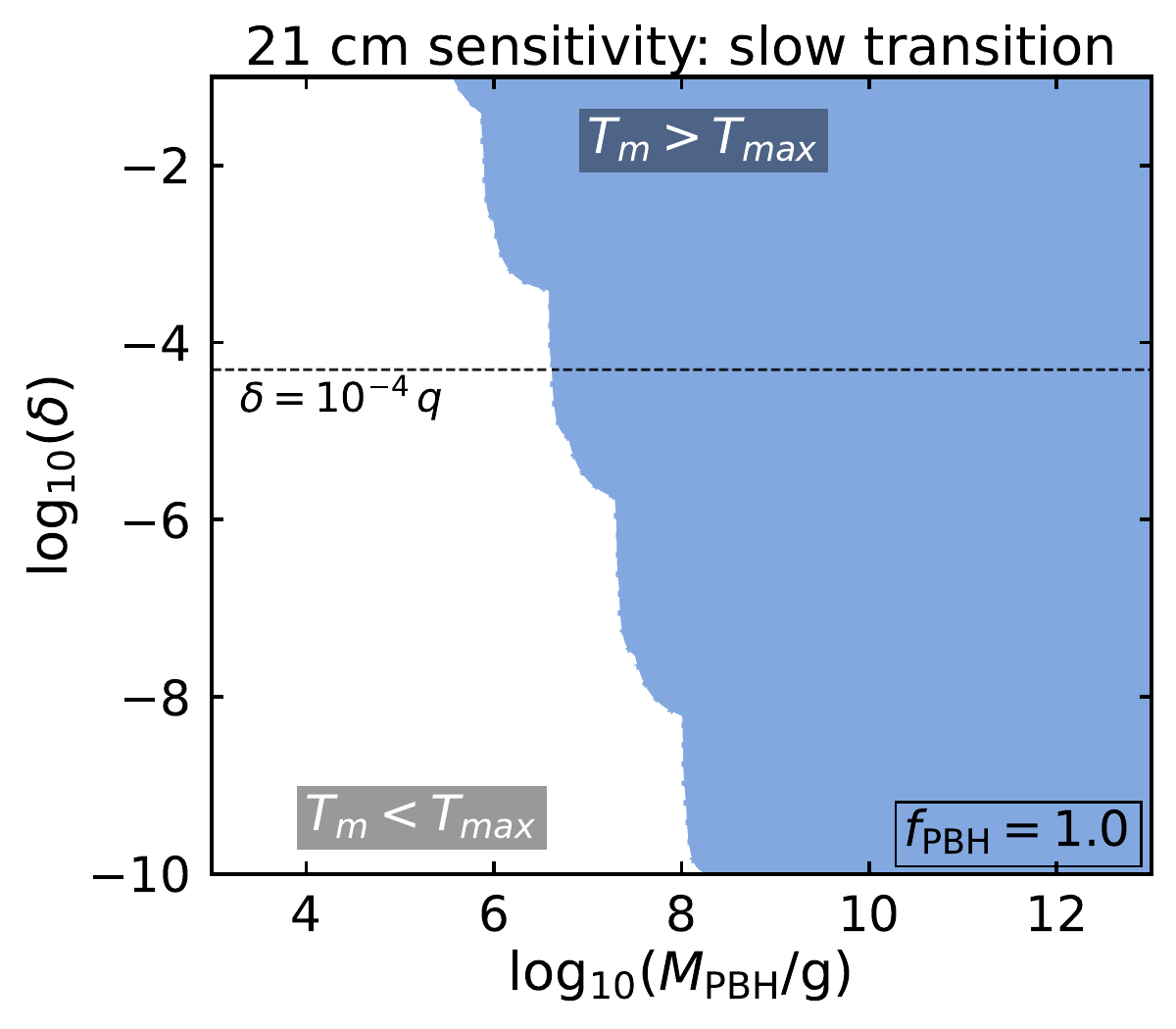}
    \caption{21 cm sensitivity for PBHs in
    fast and slow transition memory burden phase at $z=17$, shown for fixed abundance $f_{\rm PBH}=1$ . The shaded areas correspond to $T_m(z=17) > T_{\max}\,(=9.41\, {\rm K})$, with $T_m$ the predicted matter temperature.
    \emph{Left:} Fast-transition case (controlled by $k$), where the evaporation rate changes sharply. The suppression factor scales as $\mathcal{K}\propto M^{-2k}$, leading to strong suppression of Hawking emission for large $M_{\rm PBH}$. Hence, only lighter PBHs and small values of $k$ produce sufficient heating to be excluded by 21\,cm data.
    \emph{Right:} Slow transition case with a width $\delta$ of transition, where the evaporation rate evolves gradually over a finite interval. The suppression becomes weaker for larger PBH masses- due to the longer semiclassical lifetime $\tau_{\rm sc}\propto M^{3}$- high mass PBHs retain more luminosity and hence can heat the IGM efficiently. The corresponding sensitivity region, therefore, extends to larger masses. The dashed line represents $\delta = 10^{-4}q$ for $q=\frac{1}{2}$, below which the transition is effectively instantaneous.}
    \label{fig:21cmsensitivity}
\end{figure}

By computing $T_m(z=17)$ as a function of the memory burden parameters ($\delta$, $k$) and PBH properties (mass $M_{\rm PBH}$, fraction $f_{\rm PBH}$), we map out regions of parameter space compatible with 21 cm bound. 
Regions satisfying $T_m < T_{\rm max}$ remain allowed, while those with $T_m > T_{\rm max}$ are ruled out. Fig.~\ref{fig:21cmsensitivity} compares the 21 cm sensitivity to memory–burdened PBHs in the fast- and slow-transition regimes. The left panel of fig.~\ref{fig:21cmsensitivity}  shows the fast-transition scenario, in which the evaporation rate is instead abruptly reduced by a fixed entropy-suppressed factor $\mathcal{K}\!\sim\!S^{-k}$ once the black hole reaches the memory-burden threshold. The shaded region marks where PBH heating drives $T_m$ above $T_{\rm max}$. 
Because the transition is nearly instantaneous, the heating history is much shorter, and the sensitivity region becomes correspondingly narrower: only very light PBHs with $M_{\rm PBH}\lesssim 10^{7}$\,g are constrained, and only for relatively small suppression exponents $k\sim 1$--$1.5$. For larger $k$, the suppression factor $S^{-k}$ becomes so efficient that the PBH heating is negligible, causing the blue shaded region to vanish rapidly with increasing $k$. 

The right panel in fig.~\ref{fig:21cmsensitivity} corresponds to the case of slow transition, where the suppression factor evolves gradually according to $\mathcal{K}(t)\!\sim\!\delta\, \tau_{\rm sc}/(2t)$ during the transition epoch. Unlike the fast transition case, this produces a broad sensitivity region in the $(\log_{10} M_{\rm PBH}, \log_{10}\delta)$ plane. Here again, the blue-shaded area (labeled $T_m > T_{\rm max}$) predicts that the memory-burden-modified heating of the gas is sufficient to raise the gas temperature above the observational upper limit at $z\simeq 17$. 
The structure of the shaded region reflects the fact that even smaller values of $\delta$ like $\delta\sim 10^{-5}$ can produce appreciable heating for sufficiently light PBHs of mass $M_{\rm PBH}\sim 10^{7}$\,g. This is mainly because the slow transition allows  a long-lived energy injection history that scales  as $t^{-1}$. Consequently, the sensitivity region extends over nearly nine orders of magnitude in $\delta$ and covers  PBH masses from $\sim 10^{5.5}$\,g up to $\sim 10^{13}$\,g, with the boundary becoming shallower at larger $M_{\rm PBH}$ where the semiclassical lifetime $\tau_{\rm sc}$ approaches the cosmic age of the universe.
In this work, we  extend the scan down to $\delta= 10^{-10}$,  
and emphasize that such regions should be interpreted as
phenomenological extensions beyond the regime directly supported
by current microscopic derivations.
In particular, for sufficiently small $\delta$ 
and fixed $q=1/2$, the
transition becomes effectively instantaneous, and the evolution
approaches that of the fast-transition scenario. We therefore treat
these cases as providing a continuous interpolation toward the
fast-transition limit, rather than as distinct slow-transition
realizations.

\begin{figure}
    \centering
    \includegraphics[width=0.48\linewidth]{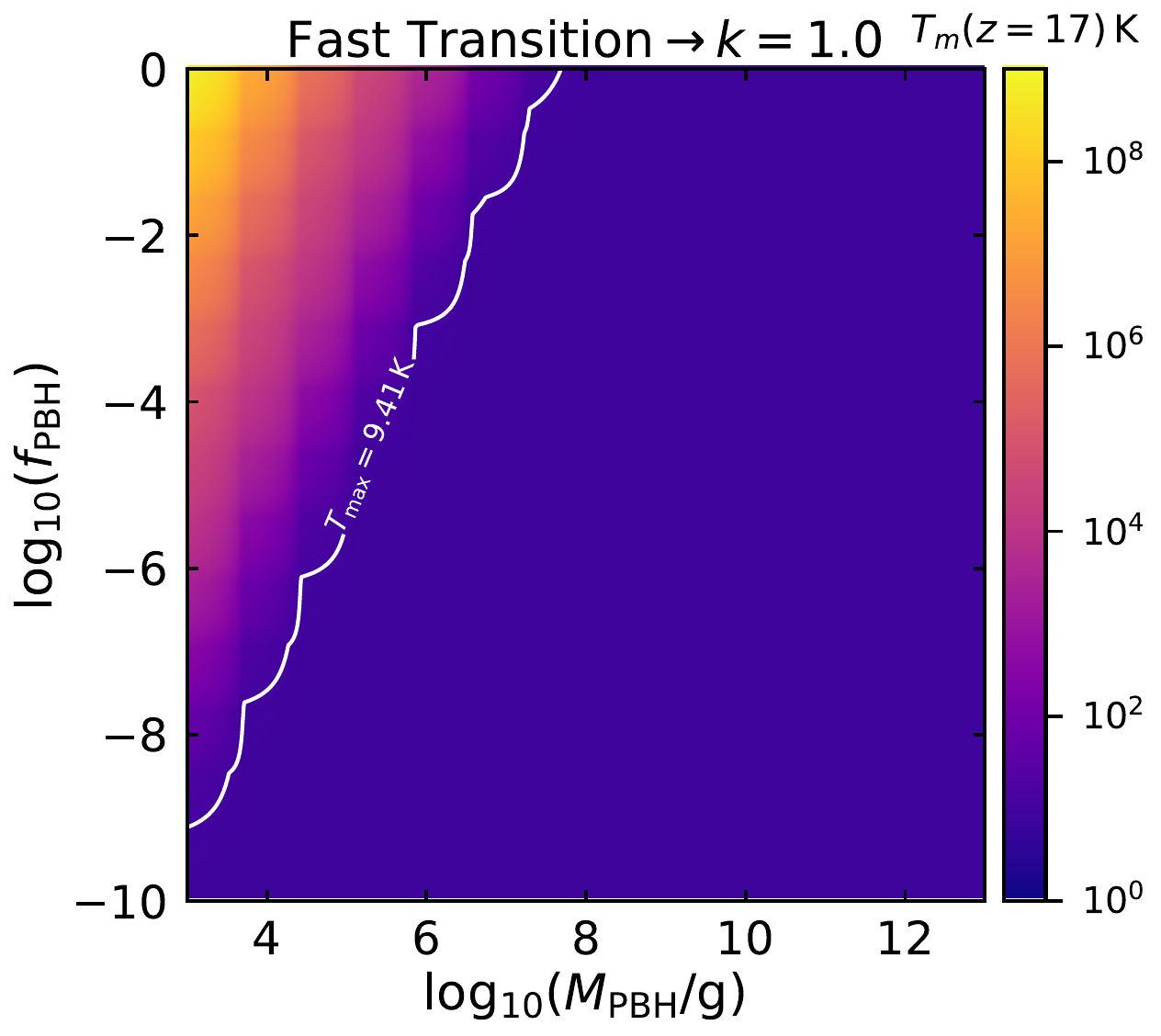}
    \includegraphics[width=0.48\linewidth]{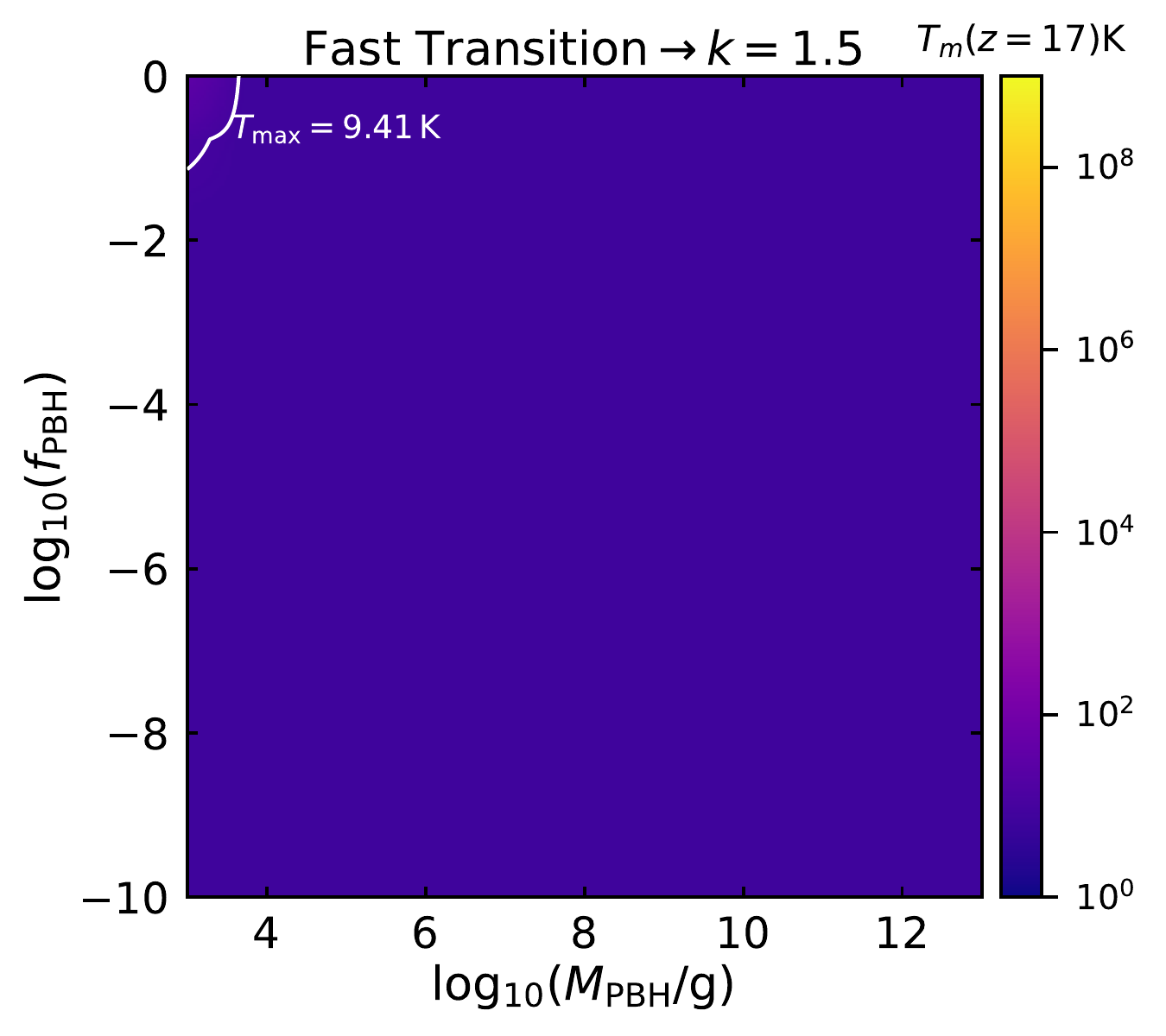}
    \caption{ Bounds on the PBH-dark matter fraction from the 21 cm signal for fast memory-burden transitions. Each panel shows the upper limit on $f_{\rm PBH}$ as a function of PBH mass $M_{\rm PBH}$ obtained by requiring the matter temperature  $T_m(z=17) \leq T_{\max}$.  The curves correspond to fast-transition memory-burden models with different suppression exponent $k$ (left: $k=1.0$, right: $k=1.5$). Larger values of $k$ induces steeper suppression $\mathcal{K}\!\propto\! M^{-2k}$, weakening the constraints at high masses.  }
    \label{fig:fvsM_fast}
\end{figure}

Now we examine the behavior of the matter temperature $T_m(z=17)$ across the 
$(M_{\rm PBH}, f_{\rm PBH})$ parameter space for two representative choices of the 
fast–transition suppression index, $k=1.0$ and $k=1.5$ in fig.~\ref{fig:fvsM_fast}. The left panel shows the $k=1$ case, where the memory burden suppression factor is relatively mild, $\mathcal{K}\sim S^{-k}$, allowing a significant fraction of Hawking emission to survive and heat the intergalactic medium. This results in a broad region of increased matter temperature at low PBH masses, $M_{\rm PBH}\sim10^{3}$--$10^{5}\,\mathrm{g}$,  for abundances $f_{\rm PBH}\gtrsim10^{-6}$. This generates a broad region in the plane where 
$T_m(z{=}17)$ exceeds the observational limit $T_{\rm max}=9.41\,\mathrm{K}$, indicated by the white 
contour. The color gradient illustrates the rapid decrease of $T_m$ with increasing $M_{\rm PBH}$ or decreasing $f_{\rm PBH}$,  reflecting how efficiently early-time heating is quenched once the PBH population is diluted or the evaporation rate is dropped.
In contrast, the right panel displays the $k=1.5$ case, in which the suppression due to memory burden is significantly stronger. Here, the spectra is quenched earlier and more sharply, leaving only a narrow corner of the parameter space capable of heating the IGM to an observable level. For almost all PBH masses and abundances shown, the temperature remains close to the minimum allowed value, and the contour $T_m=T_{\rm max}$ boils down to a tiny region near $f_{\rm PBH}\sim1$ and $M_{\rm PBH}\sim10^{3}$--$10^{4}\,\mathrm{g}$. This drastic reduction in the exclusion region reflects the sensitivity of the 21 cm heating to the strength of the memory-burden transition: increasing $k$ suppresses the energy injection deeply enough that even the light PBHs cease to produce any appreciable thermal signature. Taken together, the two panels in fig.~\ref{fig:fvsM_fast} demonstrate that fast transitions with $k\gtrsim1.5$  are effectively invisible to 21 cm constraints, whereas smaller $k$ values leave large, testable regions $(M_{\rm PBH}, f_{\rm PBH})$ space.

\begin{figure}
    \centering
    \includegraphics[width=0.48\linewidth]{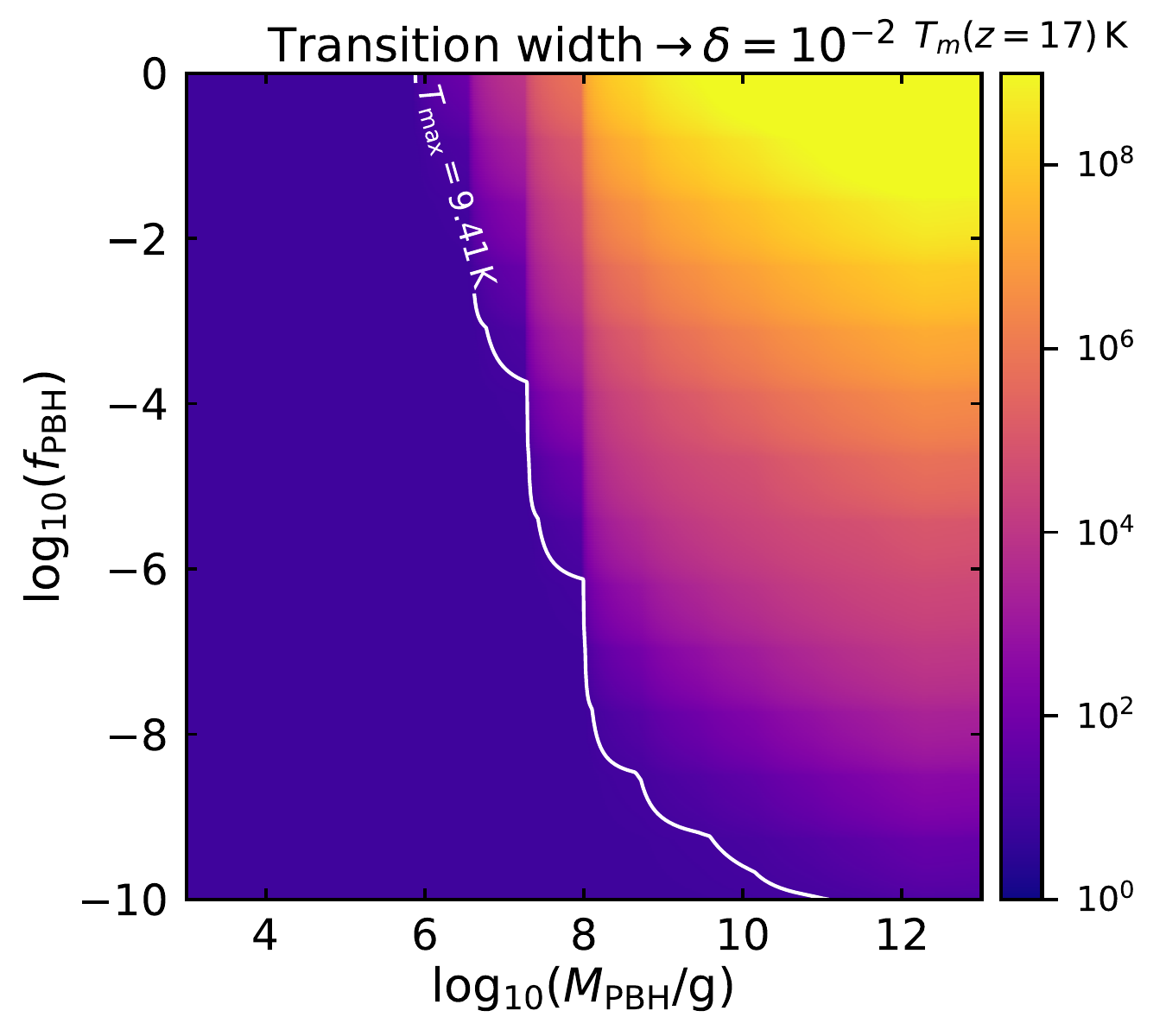}
    \includegraphics[width=0.48\linewidth]{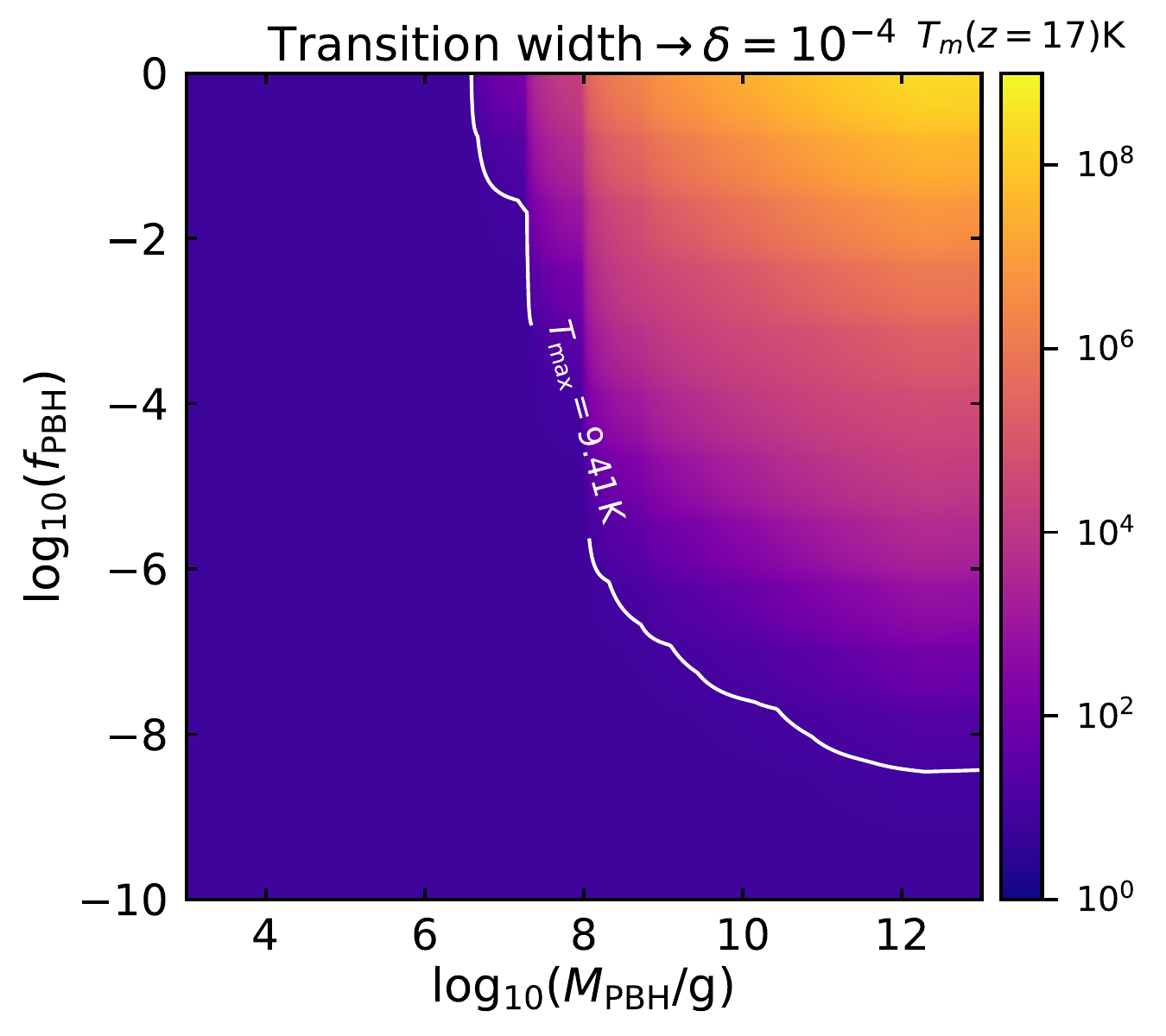}
    \caption{Bounds on the PBH-dark matter fraction from the 21 cm signal for slow memory-burden transitions. Each panel shows the upper limit on $f_{\rm PBH}$ as a function of PBH mass $M_{\rm PBH}$ obtained by requiring the matter temperature  $T_m(z=17) \leq T_{\max}$. The two panels correspond to slow-transitions with different transition widths $\delta$ (left: $\delta=10^{-2}$, right: $\delta=10^{-10}$). Smaller $\delta$ produces a sharper memory-burden suppression, reducing the emission spectra more strongly and weakening the 21\,cm constraint at high masses. Larger $\delta$ yields a more gradual transition, allowing PBHs to remain luminous for longer and tightening the resulting limits on $f_{\rm PBH}$ across a wider mass range. }
    \label{fig:fvsM_slow}
\end{figure}

An even more striking contrast emerges when we turn to the slow-transition regime, shown in \cref{fig:fvsM_slow}. For a finite transition width, $\delta=10^{-4}$, the onset of the memory-burden effect is no longer effectively instantaneous, and PBHs can continue to radiate for a non-negligible period before the Hawking flux becomes strongly suppressed. As shown in the right-panel temperature map, this leads to appreciable heating over a sizeable region of the $(M_{\rm PBH}, f_{\rm PBH})$ plane. The $T_m=T_{\rm max}$ contour extends from relatively light PBHs with $M_{\rm PBH}\sim 10^{7}\,\mathrm{g}$ at large PBH fractions and bends down to much smaller abundances, reaching $f_{\rm PBH}\sim 10^{-8}$--$10^{-9}$ for heavier masses around $M_{\rm PBH}\sim 10^{11}$--$10^{12}\,\mathrm{g}$. Thus, even for a narrow but finite slow transition, the prolonged energy injection can raise $T_m(z=17)$ above the adopted 21\,cm sensitivity threshold in a broad part of the parameter space. The left panel demonstrates a broader transition $\delta=10^{-2}$ that leads to stronger heating across a wider region of the parameter space. The $T_m=T_{\rm max}$ contour now extends to lower PBH masses, beginning around $M_{\rm PBH}\sim 10^{6}$--$10^{7}\,\mathrm{g}$ at large PBH fractions, and reaches $f_{\rm PBH}\sim 10^{-8}$--$10^{-9}$ already for $M_{\rm PBH}\sim 10^{8.5}$--$10^{9.5}\,\mathrm{g}$. For still heavier PBHs, the contour moves to even smaller abundances. In this broad-transition regime, $T_m(z=17)$ exceeds the adopted 21\,cm threshold over most of the region with $f_{\rm PBH}\gtrsim 10^{-6}$ and $M_{\rm PBH}\gtrsim 10^{8}\,\mathrm{g}$. In summary, \cref{fig:fvsM_slow} demonstrates that, unlike the fast-transition case, where increasing $k$ rapidly erases the heating signatures in the 21\,cm signal, a slow transition with finite width $\delta$ can keep PBH evaporation active for a longer period and thereby open a larger testable region of the $(M_{\rm PBH}, f_{\rm PBH})$ plane.

\medskip
\noindent
A key insight emerging from  \cref{fig:fvsM_fast} and  \cref{fig:fvsM_slow} is that the fast- and slow-transition regimes probe different PBH mass intervals, driven by the distinct mass scalings of the corresponding suppression factors. In the fast-transition case, the suppression behaves as 
$\mathcal{K}\sim S^{-k}\propto M^{-2k}$, implying that heavier PBHs experience much stronger quenching of Hawking emission. Consequently, only the lightest PBHs, $M_{\rm PBH}\lesssim10^{6}$--$10^{7}$\,g, retain enough post-transition emission to produce observable heating. By contrast, in the slow-transition regime the suppression follows $\mathcal{K}(t)\simeq\delta\,\tau_{\rm sc}/(2t)$, and since the semiclassical lifetime scales as $\tau_{\rm sc}\propto M^{3}$, more massive PBHs remain radiatively active for longer durations. This extended emission leads to significant heating for $M_{\rm PBH}\gtrsim10^{8}$--$10^{13}$\,g.

In this work, we set $x_\alpha=0$, as the Ly-$\alpha$ coupling $x_\alpha$ depends on complex astrophysical assumptions regarding the formation of the first stars and galaxies (e.g. \cite{Mittal:2020kjs} and references therein). In our analysis, we adopt a simplified treatment that avoids the uncertainties associated with specific star-formation histories.

A nonzero $x_\alpha$ would couple the spin temperature more closely to the matter temperature. In the weak-coupling limit, where both $x_c$ and $x_\alpha$ are small, the spin temperature remains close to the CMB temperature, $T_s\to T_{\rm CMB}$. In the strong-coupling limit, where either $x_c$ or $x_\alpha$ is large, the spin temperature approaches the matter temperature, $T_s\to T_m$. Thus, for intermediate coupling, $T_s$ lies between $T_m$ and $T_{\rm CMB}$. However, the effect of Ly-$\alpha$ coupling on the inferred bound is not uniformly conservative over the full parameter space. When $T_m<T_{\rm CMB}$, stronger coupling drives $T_s$ toward the colder gas temperature and can deepen the absorption signal. In contrast, in regions where PBH heating raises $T_m>T_{\rm CMB}$, stronger coupling can drive $T_s$ above the CMB temperature and qualitatively change the 21-cm signal.
Accordingly, the bounds derived here should be interpreted as projected constraints on PBH heating under the simplified thermal-history assumptions adopted in this work, rather than as bounds after marginalizing over astrophysical parameters. More generally, if a global 21-cm absorption minimum exists during cosmic dawn, an upper limit on the gas temperature at the epoch of that minimum translates into an upper limit on the allowed PBH abundance. Varying the evaluation redshift within the representative cosmic-dawn interval $z\sim15$--$20$ leads only to mild changes in the inferred bound, supporting the interpretation that the constraint is not strongly tied to a unique fiducial redshift or coupling model.


\section{Discussion and summary}\label{sec:discussion}
\begin{figure}
    \centering
    \includegraphics[width=0.48\linewidth]{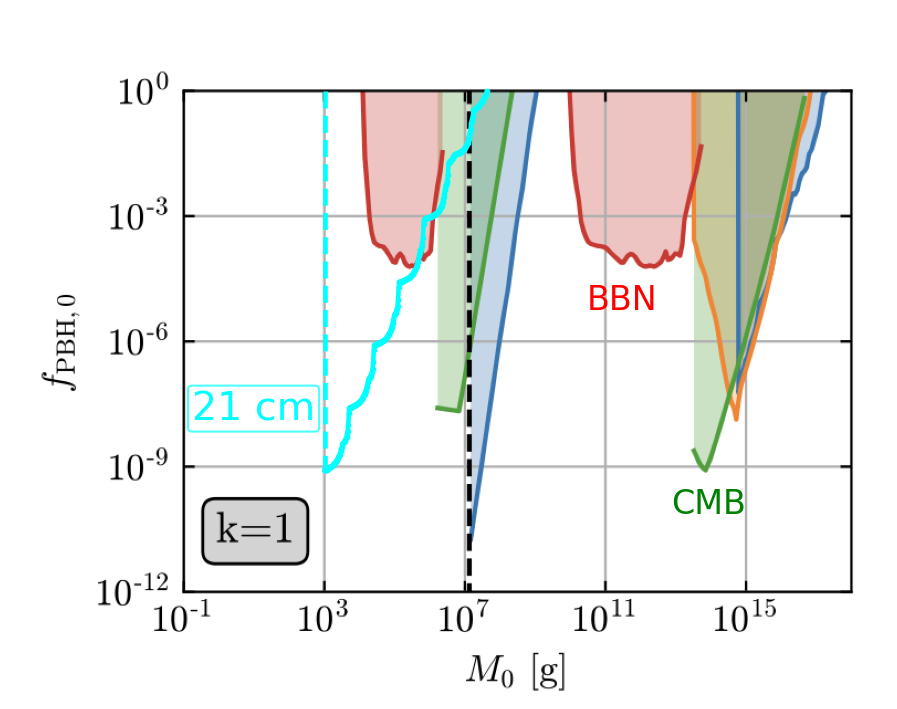}
    
    \caption{Comparison between our 21\,cm constraint on PBH-dark matter fraction for the fast transition scenario with $k=1$ (shown in cyan) and the bounds obtained from  Ref.\cite{Ft11_Chaudhuri:2025asm}. This shows the complementarity between 21 cm bounds and existing bounds from CMB (green shaded), BBN (red shaded), $\gamma$ ray emission from the 
    Galactic center (blue shaded), and extragalactic $\gamma$ ray background (orange shaded). Masses below  the dashed line are fully evaporated by the present epoch.
   For the fast transition case with $k=1$, the 21 cm bounds on the higher mass range is entirely covered by the CMB excluded region.}
    \label{fig:compare}
\end{figure}

Recent studies \cite{fastnslow, smoothtrans1_Dvali:2025ktz} have shown the memory-burden effect can substantially relax the classical Hawking constraints on light PBHs. However, the degree of suppression and the resulting viable parameter space depend sensitively on the MB transition details. 
Constraints derived from BBN require that the combination of PBH abundance and the effective transition width satisfy $[f_{\rm PBH}\,\delta ]_{\rm BBN} \lesssim 10^{-3}$ while the CMB bound requires $[f_{\rm PBH}\,\delta ]_{\rm CMB} \lesssim 10^{-9}$. In the fast transition case with $k=2$, the CMB bound extends over the full mass range $10-10^{15}\,{\rm g}$, excluding  $f_{\rm PBH}=1$. The present-day high-energy particle fluxes (gamma rays and neutrinos) impose the strongest limits at low/intermediate masses,  roughly \(f_{\rm PBH}\,\delta\lesssim 10^{-11}\) for masses near \(10^{6}\!-\!10^{8}\) g  under typical assumptions about the MB suppression exponent $k$. The global 21 cm signal at cosmic dawn provides a complementary and in many cases more stringent constraint on the slowly transitioning MB scenario because a prolonged transition injects energy at redshifts \(z\sim 10\!-\!20\) where the 21-cm brightness temperature is highly sensitive. 
Our results show that a broad transition (e.g. \(\delta\sim 10^{-2}\)) drives the allowed PBH fraction down to \(f_{\rm PBH}\lesssim 10^{-6}\) at masses \(M\sim 10^{8}\) g, whereas for a narrower transition \(\delta\sim 10^{-4}\) the 21 cm bounds weaken toward lower masses and near-unity dark matter fractions remain viable  up to \(M\sim 10^{6.5}\) g. In the fast transition case, we find that $k\gtrsim 1.5$ suppresses the heating so effectively that $T_m(z=17)$ remains below the observational upper limit $T_{\rm max}=9.41\,\mathrm{K}$ across the entire parameter space. For a milder suppression exponent $k=1$, however, we obtain a strong 21 cm exclusion at low masses: PBHs with $M_{\rm PBH}\sim 10^{3}\text{--}10^{5}\,\mathrm{g}$ are ruled out for abundances $f_{\rm PBH}\gtrsim 10^{-6}$, since their evaporation raises $T_m$ well above $T_{\rm max}$. For comparison, we overlay our bound for the fast transition case with $k=1$ onto the bounds obtained by Ref.\cite{Ft11_Chaudhuri:2025asm}, as shown in \cref{fig:compare}.

\medskip
\noindent
In summary, we have presented a comprehensive analysis of the 21\, cm constraints on primordial black holes whose evaporation is modified by the memory--burden effect. The cosmological 21 cm signal is sensitive to the long-lived, power-suppressed tail of Hawking emission that persists after the MB onset. By computing the matter temperature at $z=17$ across a wide range of PBH masses, abundances, and MB suppression parameters, we show that the strength of  21 cm constraints strongly depends on the transition dynamics. In the fast-transition case, a mild suppression ($k=1$) allows a substantial fraction of Hawking radiation to survive until late times, producing matter temperature that exceed the observational limit $T_{\rm max}(z=17)=9.41\,{\rm K}$ for PBH masses $M_{\rm PBH}\sim 10^{3}$--$10^{5}\,\mathrm{g}$ when $f_{\rm PBH}\gtrsim 10^{-6}$. A stronger suppression ($k \gtrsim 1.5$), however, quenches the evaporation so efficiently that heating becomes negligible.
In a transition regime with a relatively narrow transition width, such as $\delta=10^{-4}$, the IGM remains nearly cold only for sufficiently small PBH masses, with near-unity PBH fractions viable up to 
$M_{\rm PBH} \sim 10^{7}\,\mathrm{g}$. For larger masses, the 21\,cm bound becomes progressively stronger, reaching $f_{\rm PBH}\lesssim 10^{-6}$ at $M_{\rm PBH}\sim 10^{8}\,\mathrm{g}$ and $f_{\rm PBH}\lesssim 10^{-8}$--$10^{-9}$ for $M_{\rm PBH}\sim 10^{11}$--$10^{12}\,\mathrm{g}$. By contrast, a broader transition with $\delta = 10^{-2}$ allows the PBHs to radiate for an extended period, producing matter temperatures well above $T_{\rm max}$ over a wide mass range. In particular, we find robust 21\,cm exclusions at the level of $f_{\rm PBH}\gtrsim 10^{-8}$--$10^{-9}$ already for $M_{\rm PBH}\sim 10^{8.5}$--$10^{9.5}\,\mathrm{g}$, with the bound becoming even stronger at larger masses. In the broad-transition regime, these constraints are comparable to or stronger than the corresponding CMB constraints at the same $\delta$.

Taken together, our results highlight the importance of upcoming 21 cm observations to probe the viability of PBHs as a constituent of dark matter in the presence of novel quantum-gravitational effects such as memory burden. We emphasize that our analysis assumes spatially homogeneous energy injection and deposition from PBH evaporation. Recent studies~\cite{Sun:2023acy,Sun:2025ksr} have shown that accounting for inhomogeneous injection, including its density dependence, enhances spatial fluctuations in heating and ionization and can lead to stronger constraints, particularly from the 21-cm power spectrum. Our results based on the global (sky-averaged) signal should therefore be regarded as conservative. We are currently extending our framework to incorporate spatially inhomogeneous energy injection and to compute the corresponding 21-cm power spectrum, which will allow for a more complete assessment of the observational sensitivity to memory-burdened PBH scenarios.

\section*{Acknowledgement}
PS would like to thank the author of the \textbf{DarkHistory} package, Hongwan Liu, for useful suggestions regarding the code. PS would like to thank Akash Kumar Saha and Abhijeet Singh for suggestions and clarifications on implementing \textbf{BlackHawk} spectra to \textbf{DarkHistory} code.  The work of PS and KC was supported by the National Science and Technology Council (NSTC) of Taiwan under grant no. MOST-110-2112-M- 007-017-MY3.

\bibliography{ref}
\end{document}